\documentclass[12pt,english]{article}
\usepackage[T1]{fontenc}
\usepackage{babel}
\usepackage{graphics}

\makeatletter

\providecommand{\LyX}{L\kern-.1667em\lower.25em\hbox{Y}\kern-.125emX\@}

 \newcommand{\lyxaddress}[1]{
   \par {\raggedright #1 
   \vspace{1.4em}
   \noindent\par}
 }

\makeatother
\begin{document}

\title{Thermodynamics ``beyond'' local equilibrium}

\author{J.M.G. Vilar\( ^{\dagger } \) and J.M. Rubí\( ^{\ddagger } \)}

\maketitle

\lyxaddress{\centering \( ^{\dagger } \)Howard Hughes Medical Institute, Department
of Molecular Biology, Princeton University, Princeton, New Jersey
08544, USA}

\lyxaddress{\centering \( ^{\ddagger } \)Departament de F\'{\i}sica Fonamental,
Facultat de F\'{\i}sica, Universitat de Barcelona, Diagonal 647,
E-08028 Barcelona, Spain}

\begin{abstract}
Nonequilibrium thermodynamics has shown its applicability in a wide
variety of different situations pertaining to fields such as physics,
chemistry, biology, and engineering. As successful as it is, however,
its current formulation considers only systems close to equilibrium
---those satisfying the so-called local equilibrium hypothesis. Here
we show that diffusion processes that occur far away from equilibrium
can be viewed as at local equilibrium in a space that includes all
the relevant variables in addition to the spatial coordinate. In this
way, nonequilibrium thermodynamics can be used and the difficulties
and ambiguities associated with the lack of a thermodynamic description
disappear. We analyze explicitly the inertial effects in diffusion
and outline how the main ideas can be applied to other situations.
\end{abstract}

\newpage

Concepts of everyday use like energy, heat, and temperature have acquired
a precise meaning after the development of thermodynamics. Thermodynamics
provides us with the basis for understanding how heat and work are
related and with the rules that the macroscopic properties of systems
at equilibrium follow~{[}\ref{Reiss}{]}. Outside equilibrium, most
of those rules do not apply and the aforementioned quantities cannot
unambiguously be defined. There is, however, a natural extension of
thermodynamics to systems away from but close to equilibrium. It is
based on the local equilibrium hypothesis, which assumes that a system
can be viewed as formed of subsystems where the rules of equilibrium
thermodynamics apply. Due to the usual disparity between macroscopic
and microscopic scales, most systems fall into this category. This
is the case of, for instance, the heat transfer from a flame, the
flow through a pipe, or the electrical conduction in a wire. Nonequilibrium
thermodynamics then extracts the general features, providing laws
such as Fourier\'{ }s, Fick\'{ }s, and Ohm\'{ }s, which do not
depend on the detailed microscopic nature of the system~{[}\ref{DM}{]}.

In contrast, there are other situations where the local equilibrium
hypothesis does not hold. Many examples are present in the relaxation
of glasses and polymers~{[}\ref{Glases},\ref{Sitges},\ref{Polymers}{]},
in the flow of granular media~{[}\ref{Granular}{]}, and in the dynamics
of colloids~{[}\ref{Colloids}{]}. The main characteristic of such
systems is the similarity between microscopic and macroscopic scales,
which usually involve internal variables with {}``slow'' relaxation
times. The so-called inertial effects in diffusion processes are perhaps
the simplest and most illustrative example. In this case, the relaxation
of the velocity distribution and changes in density occur at the same
time scale. Therefore, local equilibrium is never reached. Here we
show how nonequilibrium thermodynamics, as already established in
the sixties~{[}\ref{P},\ref{DM}{]}, can be applied to this situation. 

Nonequilibrium thermodynamics~{[}\ref{DM}{]} assumes that the definition
of entropy \( S \) can be extended to systems close to equilibrium.
Therefore, entropy changes are given by the Gibbs equation: \emph{\begin{equation}
\label{dibbsxnew}
TdS=dE+pdV-\mu dN\; ,
\end{equation}
}where the thermodynamic extensive variables are the internal energy
\( E \), the volume \( V \), and the number of particles \( N \)
of the system. The intensive variables (temperature \( T \), pressure
\( p \), and chemical potential \( \mu  \)) are functions of the
extensive variables. Local equilibrium means that the Gibbs equation
holds for a small region of the space and for changes in the variables
that are actually not infinitely slow. Therefore, the internal state
of the system has to relax to equilibrium faster than variables change.
In this way, all variables retain their usual meanings and the functional
dependence between intensive and extensive variables is the same as
in equilibrium. 

Following this approach, nonequilibrium thermodynamics has been applied
to study diffusion processes. The simplest case takes place in one
dimension at constant temperature, internal energy, and volume. In
this case, from Eq.~(\ref{dibbsxnew}) we obtain a Gibbs equation
that depends only on the density and the spatial coordinate \( x \):
\emph{\begin{equation}
\label{gibbsx}
Tds(x)=-\mu (n,x)dn(x)\; .
\end{equation}
}Here \( s \) is the entropy per unit volume and \( n \) is the
density. The chemical potential has the same form as in equilibrium.
For instance, for an ideal system ---formed of non-interacting particles---
it is proportional to the logarithm of the density plus terms that
do not depend on the density~{[}\ref{DM}{]}. Notice that these terms
can include thermodynamic variables such as temperature and internal
energy, and also the spatial coordinate. In the case of non-interacting
Brownian particles, its explicit expression is \begin{equation}
\mu =\frac{k_{B}T}{m}\ln n+C(x)\; ,
\end{equation}
where \( m \) is the mass of the particles, \( k_{B} \) the Boltzmann
constant, and \( C(x) \) a function that takes into account possible
spatial inhomogeneities. The dynamics of \( n \) is restricted by
the mass conservation law and therefore follows \begin{equation}
\frac{\partial n}{\partial t}=-\frac{\partial J}{\partial x}\; ,
\end{equation}
 with \( J \) being the flux of mass. An additional assumption of
nonequilibrium thermodynamics is that this flux is given by\begin{equation}
J=-L\frac{\partial \mu }{\partial x}\; ,
\end{equation}
 where \( L \) is the phenomenological coefficient. From this, we
obtain the usual diffusion equation \begin{equation}
\frac{\partial n}{\partial t}=\frac{\partial }{\partial x}\left( D\frac{\partial n}{\partial x}\right) \; ,
\end{equation}
 with the diffusion coefficient \( D\equiv L(\partial \mu /\partial n) \). 

When inertial effects are present, changes in density occur at a time
scale comparable with the time the velocities of the particles need
to relax to equilibrium. The Gibbs equation as stated in Eq.~(\ref{gibbsx})
is no longer valid because local equilibrium is never reached. The
entropy production depends also on the particular form of the velocity
distribution. Both the spatial coordinate, \( x \), and velocity
coordinate, \( v \), are needed to completely specify the state of
the system. Therefore, we consider that local quantities are function
of both coordinates. If the system is coupled to other degrees of
freedom that relax faster than the velocity and density, a thermodynamic
description is still possible. For instance, this is the case of Brownian
particles, where the host fluid provides these thermodynamic degrees
of freedom. Thus, we consider that diffusion takes place in a two-dimensional
space (\( x,v \)) instead of in the original one-dimensional space
(\( x \)). In this case, the chemical potential for an ideal system
(e.g., non-interacting Brownian particles) is given by

\begin{equation}
\mu =\frac{k_{B}T}{m}\ln n(x,v)+C(x,v)\; ,
\end{equation}
where \( C(x,v) \) is a function that does not depend on the density~{[}\ref{DM}{]}.
The form of this function can be obtained by realizing that at equilibrium
the chemical potential is equal to an arbitrary constant. If we set
this constant equal to zero, we obtain \begin{equation}
\mu =\frac{k_{B}T}{m}\ln n+\frac{1}{2}v^{2}\; .
\end{equation}
 Therefore, the Gibbs equation is now\emph{\begin{equation}
Tds(x,v)=-\left( \frac{k_{B}T}{m}\ln n(x,v)+\frac{1}{2}v^{2}\right) dn(x,v)\; ,
\end{equation}
}The idea of applying the rules of thermodynamics in an internal space
was already proposed by Prigogine and Mazur~{[}\ref{PM}{]} and has
been used in several situations~{[}\ref{DM}, \ref{CT}{]}. In all
of them, however, there was no thermodynamic coupling of these internal
degrees of freedom with the spatial coordinate. This is precisely
the situation we are considering here. 

In the (\( x,v \))--space the mass conservation law is \begin{equation}
\label{cnxy}
\frac{\partial n}{\partial t}=-\frac{\partial J_{x}}{\partial x}-\frac{\partial J_{v}}{\partial v}\; .
\end{equation}
 Following the standard thermodynamic approach, the flux of mass is
given by\begin{equation}
\label{jx}
J_{x}=-L_{xx}\frac{\partial \mu }{\partial x}-L_{xv}\frac{\partial \mu }{\partial v}\; ,
\end{equation}
\begin{equation}
\label{jy}
J_{v}=-L_{vx}\frac{\partial \mu }{\partial x}-L_{vv}\frac{\partial \mu }{\partial v}\; ,
\end{equation}
where \( L_{ij} \), with \( i,j=\{x,v\} \), are the phenomenological
coefficients. There are some restrictions on the values \( L_{ij} \)
can take. Since the system is at local equilibrium in the (\( x,v \))-space,
Onsager relations imply that \( L_{xv}=-L_{vx} \). In addition, the
flux of mass in real space, \( \tilde{J}_{x}(x)\equiv \int _{-\infty }^{\infty }v\, n(x,v)dv \),
has to be recovered from the flux in the (\( x,v \))-space by contracting
the velocity coordinate: \( \tilde{J}_{x}(x)=\int _{-\infty }^{\infty }J_{x}(x,v)dv \).
Therefore,\begin{equation}
\int _{-\infty }^{\infty }v\, n\, dv=-\int _{-\infty }^{\infty }\left( L_{xx}\frac{k_{B}T}{m}\frac{1}{n}\frac{\partial n}{\partial x}+L_{xv}\frac{k_{B}T}{m}\frac{1}{n}\frac{\partial n}{\partial v}+L_{xv}v\right) dv\; .
\end{equation}
 Since \( n(x,v) \) can take any arbitrary form, the last equality
holds if and only if \( L_{xx}=0 \) and \( L_{xv}=-n \). Thus, the
only undetermined coefficient is \( L_{vv} \), which can depend explicitly
on \( n \), \( x \), and \( v \). 

Previous equations can be rewritten in a more familiar form by identifying
the phenomenological coefficients with macroscopic quantities. In
this way, with \( L_{vv}=n/\tau  \), the fluxes read \begin{equation}
\label{flx}
J_{x}=\left( v+\frac{D}{\tau }\frac{\partial }{\partial v}\right) n\; ,
\end{equation}
 \begin{equation}
\label{flv}
J_{v}=-\left( \frac{D}{\tau }\frac{\partial }{\partial x}+\frac{v}{\tau }+\frac{D}{\tau ^{2}}\frac{\partial }{\partial v}\right) n\; ,
\end{equation}
where \( D\equiv \frac{k_{B}T}{m}\tau  \) and \( \tau  \) are the
diffusion coefficient and the velocity relaxation time, respectively.
The equation for the density is given by\begin{equation}
\label{FP}
\frac{\partial n}{\partial t}=-\frac{\partial }{\partial x}vn+\frac{\partial }{\partial v}\left( \frac{v}{\tau }+\frac{D}{\tau ^{2}}\frac{\partial }{\partial v}\right) n\; .
\end{equation}
 This kinetic equation is equivalent to the Fokker-Planck equation
for a Brownian particle with inertia since, in an ideal system, the
density is proportional to the probability density; i.e. \( n(x,v)=mNP(x,v) \),
where \( P(x,v) \) is the probability density for a particle to be
at \( x \) with velocity \( v \), and \( N \) is the number of
particles of the system. The resulting Fokker-Planck equation could
have also been derived by following standard techniques of stochastic
processes~{[}\ref{VK}{]} or kinetic theory~{[}\ref{TC}{]}, which
are among the microscopic statistical theories for studying nonequilibrium
phenomena. 

The approach we have followed, however, explicitly illustrates how
thermodynamic concepts can be transferred from equilibrium, through
local equilibrium, to far from equilibrium situations. The condition
of equilibrium is characterized by the absence of dissipative fluxes
(\( J_{x}=0 \) and \( J_{v}=0 \)). Therefore, from Eq.~(\ref{flx})
we obtain that the velocity distribution is Gaussian with variance
proportional to the temperature. If deviations from equilibrium are
small (\( J_{x}\neq 0 \) and \( J_{v}=0 \)), the local equilibrium
hypothesis holds. This is the domain of validity of Fick's law, \begin{equation}
J_{x}=-D\frac{\partial n}{\partial x}\; ,
\end{equation}
which is obtained directly from the equations for the fluxes. In this
case, the distribution of velocities is still Gaussian but now centered
at a non-zero average velocity and the variance of the distribution
is related to the temperature in the same way as in equilibrium. Beyond
local equilibrium (\( J_{x}\neq 0 \) and \( J_{v}\neq 0 \)), the
velocity distribution can take any arbitrary form, from which there
is no clear way to assign a temperature. There is, however, a well
defined temperature \( T \): that of local equilibrium in the (\( x,v \))--space. 

\begin{figure}[h]
{\centering \resizebox*{8cm}{!}{\includegraphics{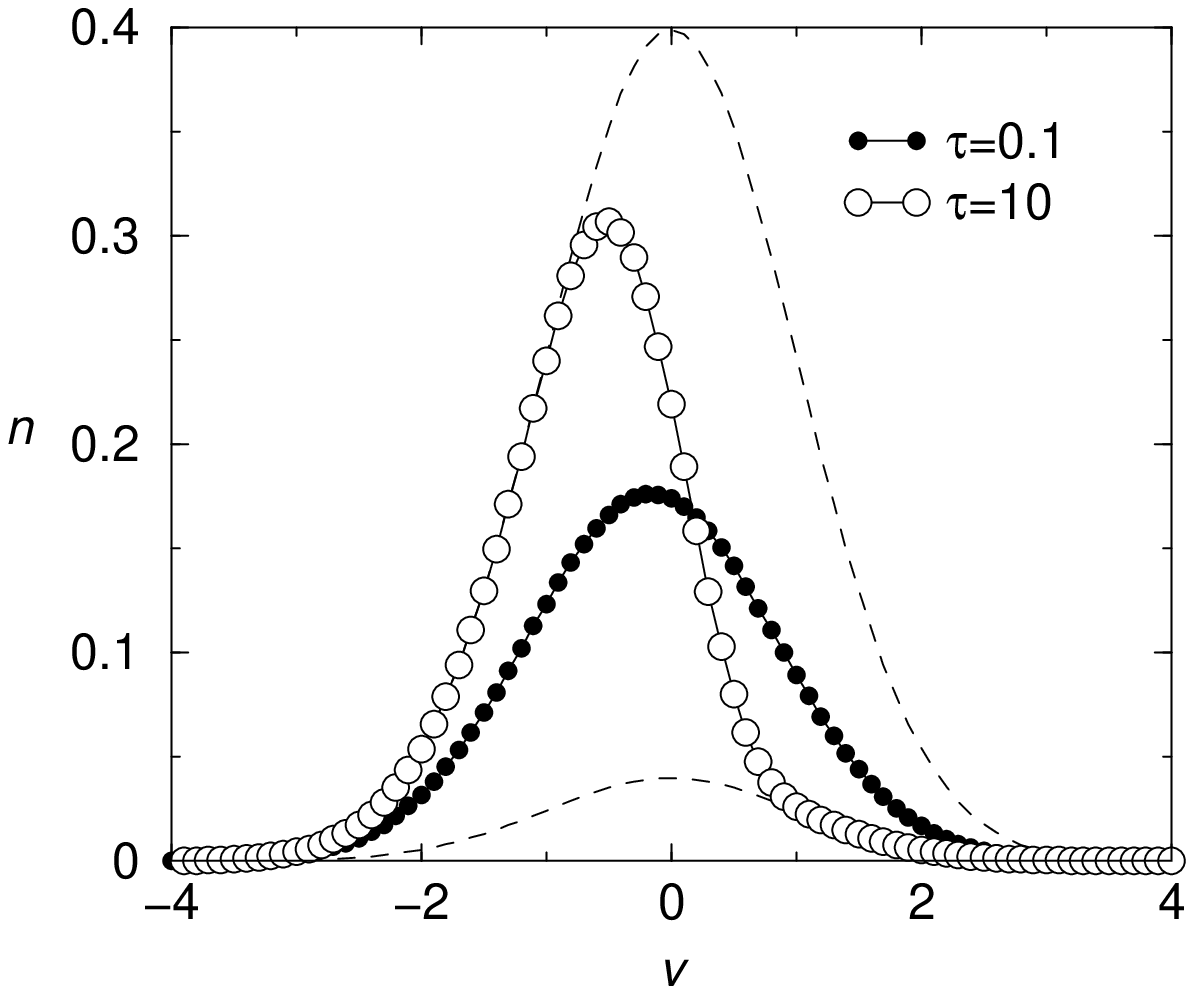}} \par}

\caption{\label{Fig:distributions} Velocity profiles obtained from Eq.~(\ref{FP})
when a density gradient is applied. The solution has been obtained
through a standard numerical algorithm following a first order upwind
discretization scheme~{[}\ref{Ames}{]}. The system is in a rectangular
domain in the (\protect\( x,v\protect \))-space, from \protect\( x=0\protect \)
to \protect\( x=1\protect \), and from \protect\( v=-10\protect \)
to \protect\( v=10\protect \). The lower and upper dashed curves
in the figure represent the boundary conditions applied at \protect\( x=0\protect \)
and \protect\( x=1\protect \), respectively: \protect\( n(1,v)=10\, n(0,v)=(2\pi )^{-0.5}\exp (-v^{2}/2)\protect \).
Filled circles correspond to velocity profiles at \protect\( x=0.5\protect \)
for fast relaxation of the velocity coordinate (\protect\( \tau =0.1\protect \)),
whereas empty circles correspond to slow relaxation (\protect\( \tau =10\protect \)).
In both cases \protect\( D/\tau \equiv k_{B}T/m=1\protect \). All
values are given in arbitrary units. }
\end{figure}

In Fig.~\ref{Fig:distributions} we illustrate the concepts discussed
previously. We show the velocity profiles obtained from Eq.~(\ref{FP})
for two representative situations. For fast relaxation of the velocity
coordinate, the velocity distribution is Gaussian and centered slightly
away from zero, in accordance with local equilibrium concepts. For
slow relaxation, however, the velocity distribution loses its symmetry
(and its Gaussian form). In this case, the temperature does not give
directly the form of the distribution and one has to resort to local
equilibrium in the (\( x,v \))--space to describe the system.

It is important to emphasize that the temperature \( T \) is the
one that enters the total entropy changes and therefore the one related
to the second principle of thermodynamics. Other definitions of temperature
are possible though. To illustrate this point, let us compute the
entropy production \( \sigma  \). This quantity is obtained from
local changes in entropy, which are given not only by the production
but also by the flow: \begin{equation}
T\frac{\partial s}{\partial t}=-\mu \frac{\partial n}{\partial t}=T\left( \sigma -\frac{\partial J_{Sx}}{\partial x}-\frac{\partial J_{Sv}}{\partial v}\right) \; ,
\end{equation}
where \( \left( J_{Sx},J_{Sv}\right)  \) is the entropy flux. In
our case, the expression for the entropy production is \begin{equation}
\label{enprod}
\sigma (x,v)=\frac{n(x,v)}{T\tau }\left( v+\frac{k_{B}T}{m}\frac{\partial \ln n(x,v)}{\partial v}\right) ^{2}\; .
\end{equation}
Now, given a Gaussian velocity distribution \( n(x,v)=n_{0}(x)e^{-mv^{2}/2k_{B}\tilde{T}(x)} \),
we can easily understand the meaning of the temperature \( \tilde{T}(x) \)
defined through the variance of the distribution: it is the temperature
at which the system would be at equilibrium (\( \sigma =0 \)). The
definition of an effective temperature as that giving zero entropy
production can be extended to arbitrary velocity distributions. From
Eq.~(\ref{enprod}), we obtain\begin{equation}
\frac{1}{\tilde{T}(x,v)}=-\frac{1}{vm}\left( k_{B}\frac{\partial \ln n(x,v)}{\partial v}\right) \; .
\end{equation}
 The temperature defined in this way is formally analogous to the
equilibrium temperature since the right hand side term of the preceding
equation can be rewritten as the derivative of an entropy with respect
to an energy:\begin{equation}
\label{tempeff}
\frac{1}{\tilde{T}(x,v)}=\frac{\partial s_{c}(x,v)}{\partial e(v)}\; ,
\end{equation}
where \( s_{c}(x,v)=-\frac{k_{B}}{m}\ln n(x,v) \) and \( e(v)=\frac{1}{2}v^{2} \).
The term \( s_{c}(x,v) \) and \( e(v) \) can be viewed as the configurational
entropy and the kinetic energy per unit of mass, respectively. In
general, other definitions of effective temperature are possible.
For instance, by considering \( e\left( v-\overline{v}(x)\right)  \)
instead of \( e(v) \) in Eq.~(\ref{tempeff}), the resulting temperature
would be that of local equilibrium. In this case, however, this temperature
does not give zero entropy production but just that of the macroscopic
motion. This temperature is then the one at which, once the macroscopic
motion is disregarded, the internal configuration of the system would
be at equilibrium.

In general, since \( \tilde{T}(x,v) \) is not only a function of
\( x \) but also of \( v \), given a point in space, there is no
temperature at which the system would be at equilibrium; i.e. \( \tilde{T}(x,v)\neq \tilde{T}(x) \).
If an effective temperature at a point \( x \) were defined, it would
depend on the way the additional coordinate is eliminated. Thus, ambiguities
in far from equilibrium quantities arise when considering a lower
dimensional space than the one in which the process is actually occurring.
This is to some extent similar to what happens with effective temperatures
defined through fluctuation-dissipation theorems. In such a case,
the effective temperature can depend on the scale of observation~{[}\ref{kurchan}{]}.
It is interesting to point out that all these effective temperatures,
despite their possible analogies with the equilibrium temperature,
do not have to follow the usual thermodynamic rules since the system
is not actually at equilibrium at the temperature \( \tilde{T} \). 

The idea of increasing the dimensionality of the space were diffusion
takes place, so to include as many dimensions as non-equilibrated
degrees of freedom the system has, can also be applied to other situations.
In a general case, the additional degrees of freedom do not necessarily
correspond to the velocity. For instance, let us consider a degree
of freedom \( \Theta (x) \) that at local equilibrium enters the
Gibbs equation in the following way:\begin{equation}
Tds(x)=-\mu \, dn(x)-B\, d\Theta (x)\; ,
\end{equation}
 where \( B\equiv B(n,\Theta ,T)=T(\partial s/\partial \Theta )_{n,T} \).
In this case, one usually assumes that given \( T \) , \( n(x) \),
and \( \Theta (x) \), the function \( B \) is completely determined
through the equilibrium properties of the system. Far away from equilibrium,
we would have to consider explicitly an additional coordinate \( \theta  \),
which is related to the degree of freedom by \( \Theta (x)=\int \theta \, n(x,\theta )d\theta  \).
The corresponding Gibbs equation\begin{equation}
Tds(x,\theta )=-\mu \, dn(x,\theta )
\end{equation}
would have to take into account the dependence on the coordinate \( \theta  \)
through the chemical potential \( \mu  \). Once the Gibbs equation
has been obtained, the way to proceed would be analogous to the one
we followed for the inertial effects. For instance, some systems with
both translational and orientational degrees of freedom can be described
by the chemical potential\begin{equation}
\mu =\frac{k_{B}T}{m}\ln \frac{n(x,\theta )}{f(\theta )}+U\cos \theta 
\end{equation}
 where \( \theta  \) is now an angular coordinate, \( U\cos \theta  \)
is the orientational energy, and \( f(\theta ) \) is a function accounting
for the degeneracy of the orientational states (for rotation in three
and two dimensions, \( f(\theta )=\sin \theta  \) and \( f(\theta )=1, \)
respectively)~{[}\ref{DM}{]}. This type of systems include, among
others, liquid crystals and suspensions of rod-like particles~{[}\ref{Polymers}{]},
field-responsive suspensions~{[}\ref{jms}{]}, and polarized systems~{[}\ref{DM}{]}.
At local equilibrium some instances of \( B \) and \( \Theta  \)
are then electric field and polarization; and magnetic field and magnetization.
Beyond local equilibrium, by writing the \( (x,\theta ) \) counterpart
of Eqs.~(\ref{cnxy}), (\ref{jx}), and (\ref{jy}), one can obtain
a kinetic equation that describes the behavior of the system. This
equation includes as particular cases the Fokker-Plank equations obtained
for those systems by means of microscopic theories~{[}\ref{Polymers},\ref{BP}{]}.

Along this paper, we have been assuming ideality and locality. The
condition of ideality is that the system consists of non-interacting
particles. In this case, the chemical potential is proportional to
the logarithm of the density plus terms that do not depend on this
quantity. Non-ideality can be directly taken into account by considering
the right dependence of the thermodynamic quantities on the density
and, in general, will give rise to nonlinear partial differential
equations. A more difficult aspect to deal with is the presence of
non-local effects. In such a case, the interactions between the different
parts of the system will need of integro-differential equations to
be incorporated in the description. 

The main result of our analysis shows that in far from equilibrium
diffusion processes, local equilibrium can be recovered when all the
relevant degrees of freedom are considered at the same level as the
spatial coordinate. In the resulting extended space, thermodynamic
quantities, such as temperature and the chemical potential, admit
a well defined interpretation. The scheme we have developed may then
provide the basis for a consistent formulation of thermodynamics far
from equilibrium.

\emph{Acknowledgements}--- J.M.R was supported by DGICYT (Spain) Grant
No. PB98-1258. J.M.G.V. is an associate of the Howard Hughes Medical
Institute.

\newpage

\end{document}